\documentclass[fleqn,usenatbib]{mnras}

\usepackage{newtxtext,newtxmath}

\usepackage[T1]{fontenc}

\DeclareRobustCommand{\VAN}[3]{#2}
\let\VANthebibliography\thebibliography
\def\thebibliography{\DeclareRobustCommand{\VAN}[3]{##3}\VANthebibliography}

\usepackage{graphicx}	
\usepackage{amsmath}	
\usepackage{siunitx}
\usepackage{subcaption}
\usepackage{threeparttable}
\usepackage{float}
\usepackage{anyfontsize}
\usepackage{lscape} 
\usepackage{caption}

\DeclareCaptionLabelFormat{continued}{#1~#2 (Cont.)}
\DeclareSIUnit\angstrom{\text{Å}}
\providecommand\phantomcaption{\caption@refstepcounter\@captype}

\title{Characterization of hot populations of Melotte 66 open cluster using Swift/UVOT}

\author[Khushboo K. Rao et al.]{
Khushboo K. Rao,$^{1}$ \thanks{E-mail: p20170419@pilani.bits-pilani.ac.in} 
Kaushar Vaidya,$^{1}$ 
Manan Agarwal,$^{2}$
Anju Panthi,$^1$
Vikrant Jadhav,$^{3,4}$
\newauthor and Annapurni Subramaniam$^{3}$
\\
$^{1}$ Department of physics, Birla Institute of Technology and Science-Pilani, 333031 Rajasthan, India.\\
$^{2}$ Department of Physics and Kavli Institute for Astrophysics and Space Research, Massachusetts Institute of Technology, Cambridge, MA 02139, USA. \\
$^{3}$Indian Institute of Astrophysics, Sarjapur Road, Koramangala, Bangalore, India. \\
$^{4}$Joint Astronomy Programme and Department of Physics, Indian Institute of Science, Bangalore, India.}

\date{Accepted. Received YYY; in original form ZZZ}

\pubyear{2015}

\begin{document}
\label{firstpage}
\pagerange{\pageref{firstpage}--\pageref{lastpage}}
\maketitle

\begin{abstract}
Ultraviolet (UV) wavelength observations have made a significant contribution to our understanding of hot stellar populations of star clusters. Multi-wavelength spectral energy distributions (SEDs) of stars, including ultraviolet observations, have proven to be an excellent tool for discovering unresolved hot companions in exotic stars such as blue straggler stars (BSS), thereby providing helpful clues to constrain their formation mechanisms. Melotte 66 is a 3.4 Gyr old open cluster located at a distance of 4810 pc. We identify the cluster members by applying the ML-MOC algorithm on \textit{Gaia} EDR3 data. Based on our membership identification, we find 1162 members, including 14 BSS candidates, 2 yellow straggler candidates (YSS), and one subdwarf B candidate (sdB). We generated SEDs for 11 BSS candidates and the sdB candidate using Swift/UVOT data combined with other archival data in the optical and IR wavelengths. We discover a hot companion of one BSS candidate, BSS3, with temperature of 38000$_{-6000}^{+7000}$ K, luminosity of 2.99$_{-1.86}^{+5.47}$ L$_{\sun}$, and radius of 0.04$_{-0.005}^{+0.008}$ R$_{\sun}$. This hot companion is a likely low-mass WD with an estimated mass of 0.24 -- 0.44 M$_{\sun}$. We report one BSS candidate, BSS6, as an Algol-type eclipsing binary with a period of 0.8006 days, based on the \textit{Gaia} DR3 variability classification. We suggest that BSS3 is formed via either the Case A or Case B mass-transfer channel, whereas BSS6 is formed via the Case A mass transfer.
\end{abstract}

\section{Introduction}
It is widely accepted that all stars form in clusters \citep{Portegies2010}. Star clusters are host to both common stellar populations that can be explained by stellar evolution theory and exotic populations such as blue straggler stars \citep[BSS;][]{Bailyn1995}, cataclysmic variables \citep{Ritter2010}, millisecond pulsars \citep{Bhattacharya1991}, subdwarf stars \citep{Moehler2001,Bidin2008}, and many more that emerge from either binary system evolution or stellar interactions. BSS and subdwarf stars among these exotic populations can be identified because they populate an unexpected area of the color-magnitude diagram (CMD) or the Hertzsprung-Russel (H-R) diagram. Furthermore, their location indicates that they have a history that has influenced their temperature and luminosity. The emergence of \textit{Gaia}'s thriving era of precise astrometry \citep{GaiaMission2016} combined with the revolution in computational capabilities  \citep{Cantat2018,Agarwal2021}, enabled us to classify these stellar populations associated with their host clusters more precisely \citep{Rain2021,Geier2020,Lei2020}. These exotic populations are hotter because they are the products of the complex interactions of two or more stars. Since they are hotter, they emit a significant amount of their flux at UV wavelengths. Therefore, in the last decade UV wavelengths combined with optical to IR wavelengths have been used to construct spectral energy distributions (SEDs) of these populations to estimate their fundamental parameters as well as detect associated hot companions. For example, globular clusters NGC 5466 \citep{Sahu2019} and NGC 1851 \citep{Singh2020}, and open clusters NGC 188 \citep{Gosnell2014,Gosnell2015,Subramaniam2016,Rani2021}, NGC 2682 \citep{Sindhu2019,Jadhav2019}, and NGC 7789 \citep{Vaidya2022}. 

BSS have long been of particular interest among exotic populations because they are on the extension of hydrogen burning phase which is not possible according to the cluster age \citep{Sandage1953}. The hydrogen burning phase revives as a consequence of either collisions between unevolved main sequence stars or mass transfer/merger in primordial binaries \citep{Stryker1993S,Boffin2015}. The collisional formation channel is the most plausible for dense stellar environments, such as the cores of globular clusters, whereas formation channels involving binary systems have a few subcategories and are favoured channels for both low to highly dense environments such as open clusters, globular clusters, dwarf galaxies and Galactic fields \citep{Ferraro1993,Ferraro1995}. The classification of BSS formation mechanism via mass transfer in a binary system is done based on the location of the primary star on the CMD. Mass transfer may occur when the primary star is on the main sequence, resulting in either a single massive BSS or a binary BSS with a short period main sequence star as a companion \citep [Case A;][] {Webbink1976}. Mass transfer in a binary system may also occur as a result of Roche Lobe overflow when the primary star is on the red giant branch, i.e. RGB, resulting in a short period binary with a Helium (He) white dwarf (WD) of mass $\leq$0.45 M$_{\sun}$ as a companion \citep [Case B;] [] {McCrea1964} or when the primary star is on the asymptotic giant branch, i.e. AGB, resulting in a long period binary \citep[Case C;][]{Chen2008,Gosnell2014}.

Hot subdwarfs are core helium burners with a mass of $\sim$0.5 M$_{\sun}$ and a thin hydrogen envelope \citep{Heber1986}. They are classified into two types: subdwarf B stars (sdB), which populate the end of the extreme horizontal branch and have temperatures ranging from 20000 K to 40000 K, and subdwarf O stars, which are a mixture of post-HB, post-AGB, and post-RGB stars with temperatures higher than sdB stars \citep{Lynas2004,Heber2009}. sdB stars, like BSS, exist in single and binary systems, but binary sdB are more common than single sdB \citep{Maxted2001,Napiwotzki2004,Copperwheat2011}. Single hot sdB can form when two helium white dwarfs fuse together, whereas binary systems have red giant progenitors whose hydrogen envelope is almost entirely removed at or just before the Helium Flash \citep{Han2002,Han2003}. For a detailed review on such an interesting objects readers are referred to \citet{Heber2016}.

Melotte 66 (RA = 111.573, DEC = $-$47.685) is a well-studied open cluster, located below the Galactic plane in the Puppis constellation. \citet{Vaidya2020} recently investigated the BSS population of this cluster using the \textit{Gaia} DR2 \citep{GaiaDR22018} data. Based on the radial distribution of BSS, they inferred that Melotte 66 is an intermediate dynamical age cluster. To search for multiple stellar populations of the cluster, \citet{Carraro2014} obtained high-quality CCD UBVI photometry and high-resolution spectra of 7 red-clump stars of this cluster using the Las Campanas Observatory (LCO). They estimated the cluster distance as $\sim$4.7$^{+0.2}_{-0.1}$ kpc, the age as 3.4$\pm$0.3 Gyr, and the reddening E(B$-$V) as 0.15$\pm$0.02 mag. In order to explore the radial distribution of metals in the Galactic disk, \citet{Sestito2008} collected high-resolution UVES spectra from FLAMES@VLT of 26 RGB stars from four open clusters, including 7 RGB stars from Melotte 66. They estimated the [Fe/H] of Melotte 66 as $-$0.33$\pm$0.03. Although, Melotte 66 has received a lot of attention in the optical wavelengths, cluster's BSS population, as well as other hot populations, have not been investigated using the available UV fluxes of the sources from the Swift/UVOT data. Hence, in the present work, we study the UV population of Melotte 66 using Swift/UVOT to determine the fundamental parameters of diverse UV populations and to search for hot companions in BSS by examining their SEDs. The rest of the paper is arranged as per the following. In \S \ref{sec:data}, we give the information about the data used for the present work and membership determination of the cluster. In \S \ref{sec:CMD}, we present the CMD of the cluster. In \S \ref{sec:SEDs}, we construct and analyze the SEDs of BSS and the sdB candidates. In the end, final remarks are presented in \S \ref{sec:Final_remarks}.
\section{Data and Membership identification}
\label{sec:data}
\begin{figure*}
	\includegraphics[width=1.0\textwidth]{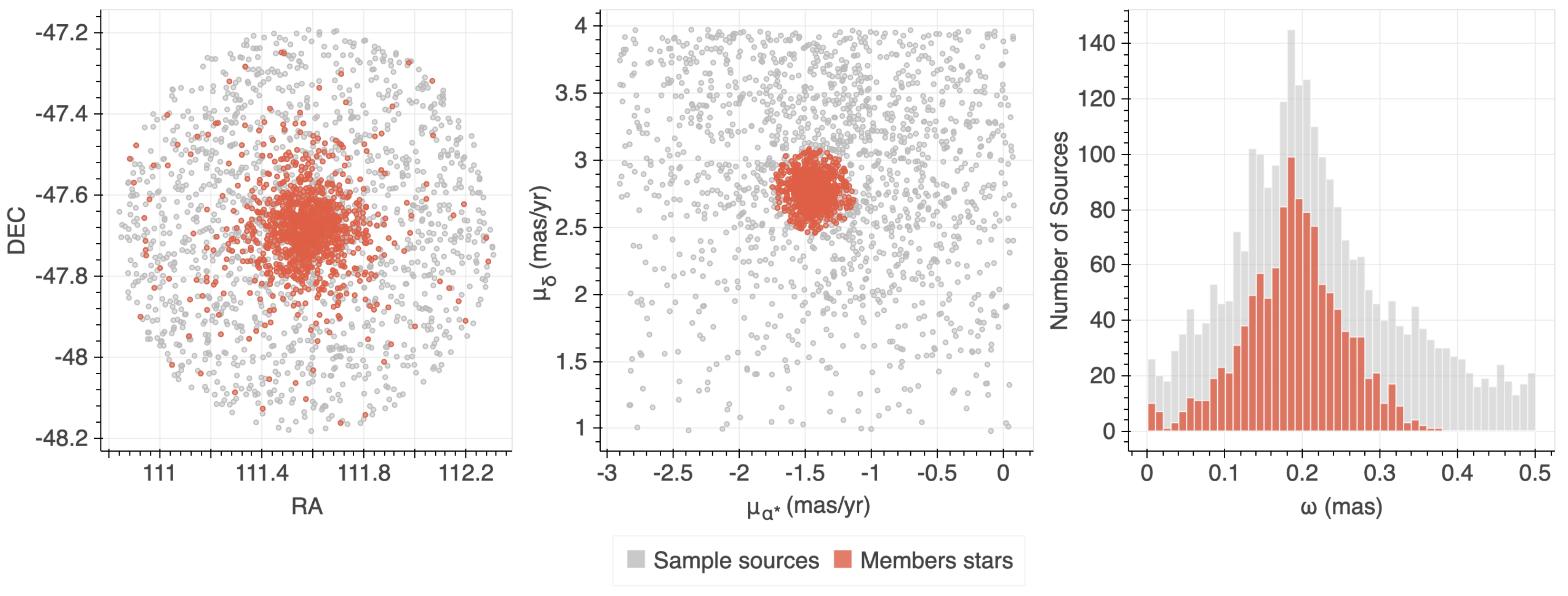}
    \caption{The spatial, proper motion and parallax distribution of \textit{\textit{sample} \textit{sources}} and cluster members identified using the ML-MOC algorithm on the \textit{Gaia} EDR3 data.}
    	\label{fig:Figure1}
\end{figure*}
In the present work, we use multi-wavelength archival data, from NUV to IR to study various stellar population of Melotte 66 open cluster.
\subsection{\textit{Gaia} EDR3 data}
To identify the cluster members, we use ML-MOC algorithm \citep{Agarwal2021} on the \textit{Gaia} EDR3 data \citep{Gaiaedr32021}. We briefly describe the procedure for identifying cluster members here, while for more details on ML-MOC, readers are referred to \citet{Agarwal2021}. The ML-MOC is a robust machine-learning based algorithm that uses k-Nearest Neighbour \citep[kNN;][]{Cover_kNN} and Gaussian mixture model \citep[GMM;][]{mclachlan200001} algorithms, which does not require any prior knowledge about a cluster. To filter out cluster members from the field stars, it uses proper motion and parallax information from \textit{Gaia} EDR3 data. 

The methodology for identifying members consists of three steps. First, the sources that fulfill the following three conditions are labelled as \textit{All} \textit{Sources}, i) the five astrometric parameters, i.e., positions, proper motions, and parallaxes along with the photometric information in all the three Gaia bands, G, G$_\mathrm{BP}$, and G$_\mathrm{RP}$ are available in the catalog, ii) the parallaxes are non-negative, and iii) the errors in G mag are smaller than 0.005 mag. kNN is then applied to the combined proper motion and parallax space of \textit{All} \textit{Sources} to eliminate probable field stars and generate \textit{sample} \textit{sources} with a higher fraction of cluster members than field stars. In the second stage, GMM is applied to the \textit{sample} \textit{sources} which then segregate cluster members from field stars based on the Gaussian distributions fitted to the \textit{sample} \textit{sources}. GMM then provides us with the mean values of the proper motions and parallax as well as assign probability to the cluster members from 0.6 -- 1.0. In the end, moderate probability sources with probabilities ranging from 0.2 -- 0.6 are included by increasing proper motion range but keeping the parallax range fix as that of the sources with membership probability > 0.8. Figure \ref{fig:Figure1} shows the spatial, proper motion, and parallax distribution of \textit{sample} \textit{sources} and cluster members. A total of 1162 sources are identified as members of Melotte 66. We also determine the cluster radius, i.e., the distance from the cluster center at which cluster members blend with field stars, as 15$\arcmin$. The estimated radius of the cluster is identical to the previous estimate of the cluster radius by \citet{Vaidya2020}. 

\subsection{Swift/UVOT data}
The Swift spacecraft carries three instruments, Ultra-Violet Optical Telescope (UVOT), X-ray telescope (XRT), and Burst Alert Telescope (BAT), essentially to allow the most thorough observations of gamma-ray bursts to date. The UVOT is a 30 cm modified Ritchey-Chretien UV/Optical Telescope co-aligned with the XRT and mounted on the common telescope platform for all instruments. It provides simultaneous optical and ultraviolet coverage (170 -- 650 nm) in a 17$\arcmin \times 17\arcmin$ field utilizing seven v, b, u, UVW1, UVM2, UVW2 and white filters \citep{Roming2005}. Melotte 66 was observed using Swift/UVOT in three UV filters, UVW1, UVM2, and UVW2, on 8, 11, 13, and 15 Jun 2011. The details of the observations are listed in Table \ref{tab:Table1}. The near ultra-violet (NUV) point source catalog of this cluster is provided by \citet{Siegel2019} and stored as level 3 data on the Swift/UVOT archive\footnote{\url{https://archive.stsci.edu/swiftuvot/search.php}}. The details of the total number of detections and counterparts of cluster members in all the three filters are given in table \ref{tab:Table1}. A total of 32 cluster members have been detected in all the three UV filters.

\subsection{Spitzer/IRAC data}
Melotte 66 was observed on 25th November 2006 under program ID 30800 in all the four channels of Spitzer/IRAC \citep{Fazio2004}, i.e., I1 (3.6 $\micron$), I2 (4.5 $\micron$), I3 (5.8 $\micron$), and I4 (8.0 $\micron$). The catalog of the cluster sources detected through the Spitzer/IRAC is provided by \citet{Spitzer2021yCat} and is available on Vizier. We found 725 members counterparts in I1 channel, 751 members counterparts in I2 channel, 319 members counterparts in I3 channel, and 166 members counterparts in I4 channel within 1$\arcsec$ tolerance. As we are particularly interested in studying UV populations, we only look for the counterparts of the 32 cluster members detected in the three UV filters of Swift/UVOT. Among the 32 sources, we find 26 counterparts in the I1 channel, 27 counterparts in the I2 channel, 19 counterparts in the I3 channel, and 20 counterparts in the I4 channel. 

\subsection{TESS data}
\label{TESS_data}
The Transiting Exoplanet Survey Satellite \citep[TESS;][]{Ricker2015} is conducting nearly all sky photometric survey in 600 -- 1000 nm wavelength range, where it obtains 30 minute cadence observation for all the sources in its FOV and 2 minute cadence observation for 200,000 -- 400,000 preselected sources. The orbit of TESS is elliptic with a period of 13.7 days, where each sector is observed for continuous 27.4 days. It transmits the data to earth when it comes on perigee, i.e, twice for each sectors. Therefore, it is sensitive to only short period exoplanets or binaries with period smaller than 13 days \citep{Stassun2018}. Melotte 66 was observed through TESS in 7, 8, 33, 34, and 35 sectors. \citet{Nardiello2020} extracted the light curves of Melotte 66 sources using the PATHOS pipeline (A PSF-based Approach to TESS High Quality data Of Stellar clusters) and made available through the STScI MAST server. They aim to extract high-quality light curves of exoplanets and variable stars of star clusters and young associations \citep{Nardiello2019,Nardiello2020}. 

\begin{table}
	\centering
	\caption{Details of the Swift/UVOT observations.}
	\label{tab:Table1}
	\begin{tabular}{ccccc} 
		\hline
		\\
		Filter & $\lambda_{\mathrm{eff}}$ & Exposure & Detections & Counterparts  \\
		 & ($\si{\angstrom}$) & (sec)  & & (ML-MOC members) \\
		\\
		\hline
		\\
		UVW2 & 2085.73 & 2095 & 586 & 71 \\
		UVM2 & 2245.78 & 2095 & 411 & 53  \\
		UVW1 & 2684.14 & 1861 & 2507 & 441 \\
		\\
		\hline
	\end{tabular}
\end{table}
\begin{figure*}
    \centering
	\begin{subfigure}[b]{0.48\textwidth}
    		\includegraphics[width=1.0\textwidth]{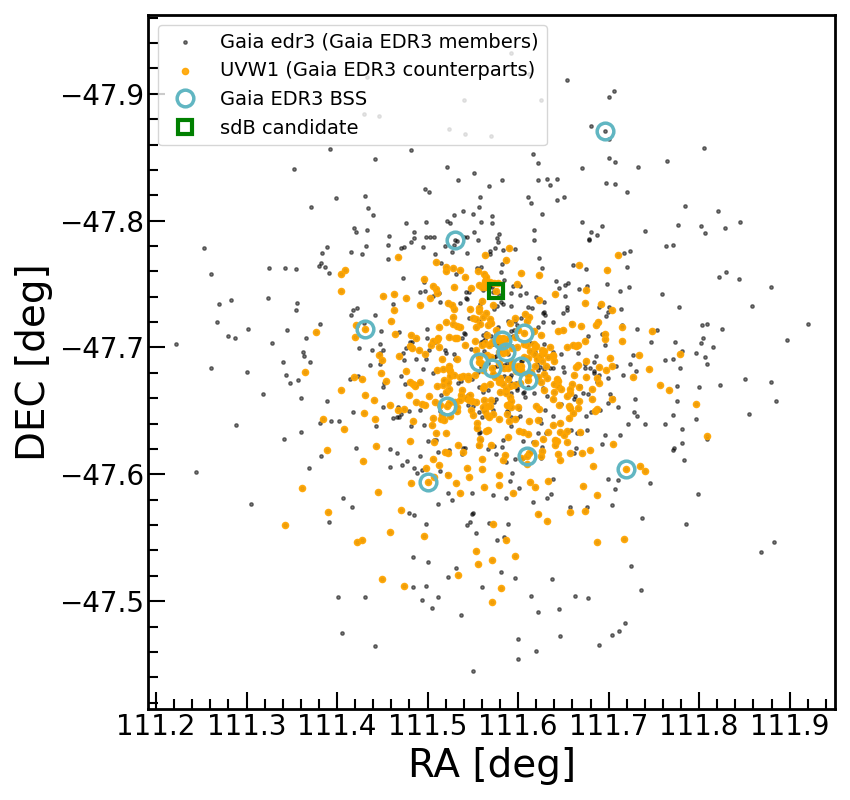}
		\caption*{}
	\end{subfigure}
	\quad 
	\begin{subfigure}[b]{0.48\textwidth}
   		\includegraphics[width=1.0\textwidth]{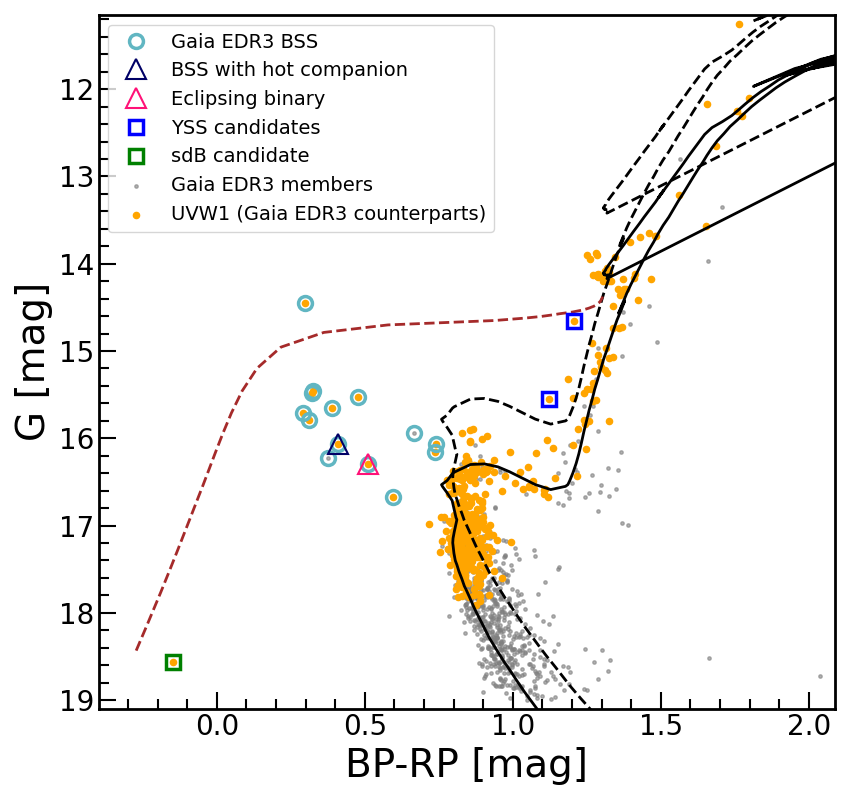}
		\caption*{}
	\end{subfigure}
    \caption{The spatial distribution (left panel) and the CMD (right panel) of cluster members identified using the ML-MOC algorithm on \textit{Gaia} EDR3 data. The cluster members are shown as grey dots, whereas the UVW1 counterparts of cluster members cross-matched within 1$\arcsec$ tolerance are shown as orange dots. The black solid line shows the plotted PARSEC isochrone. The ZAHB is shown as a brown dashed line and black dashed line shows equal-mass binary isochrone. The BSS, YSS, and sdB candidates identified in this cluster are shown as light blue open circles, blue open squares, and green open square, respectively. The dark blue open triangle shows the BSS candidate, BSS3, for which an unresolved hot companion is discovered in the current work. The magenta open triangle shows the eclipsing binary BSS (BSS6).}
    \label{fig:cmd}
\end{figure*}
\begin{figure*}
    \centering
	\begin{subfigure}[b]{0.48\textwidth}
    		\includegraphics[width=1.0\textwidth]{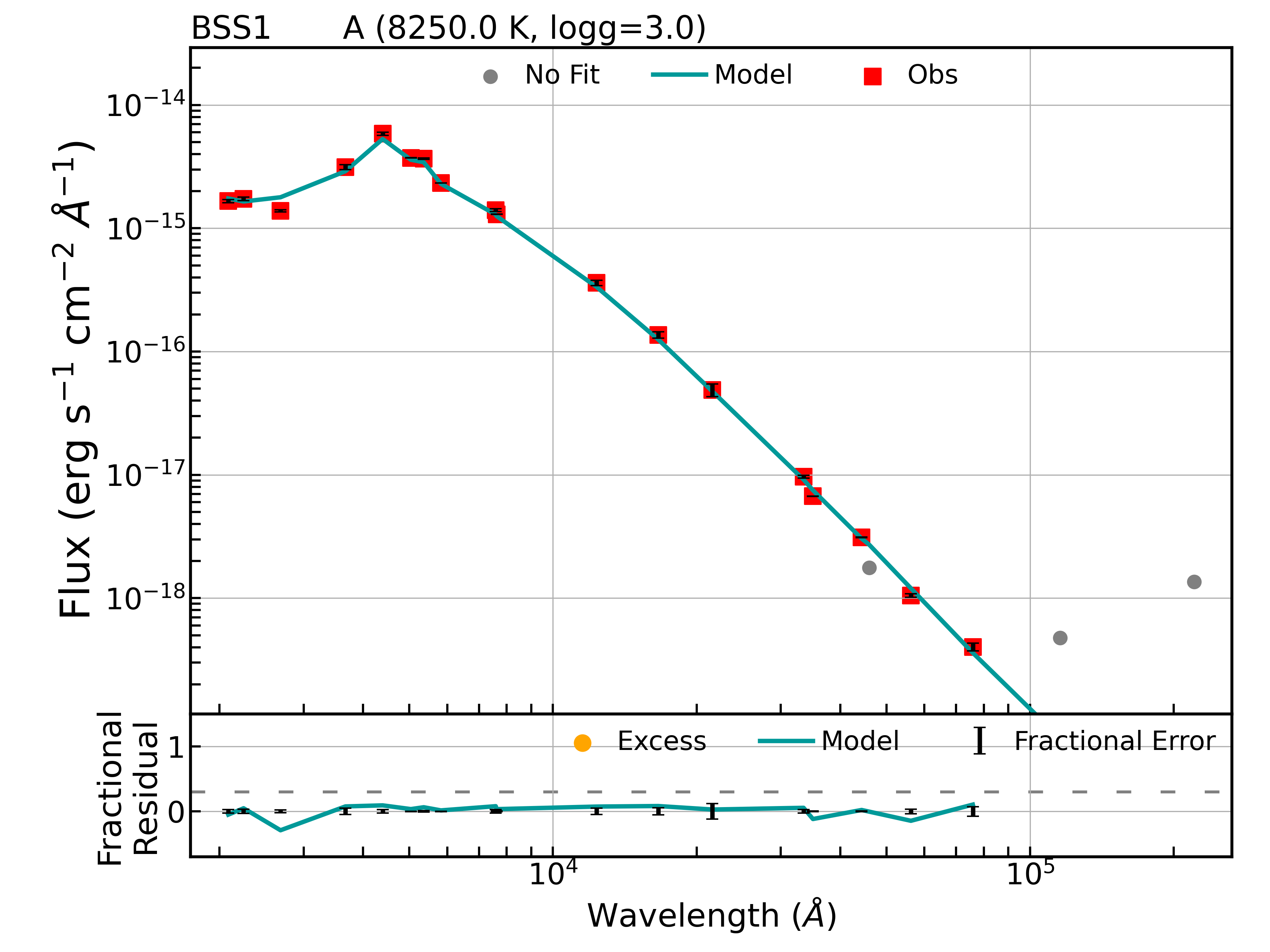}
		\caption*{}
	\end{subfigure}
	\quad 
	\begin{subfigure}[b]{0.48\textwidth}
   		\includegraphics[width=1.0\textwidth]{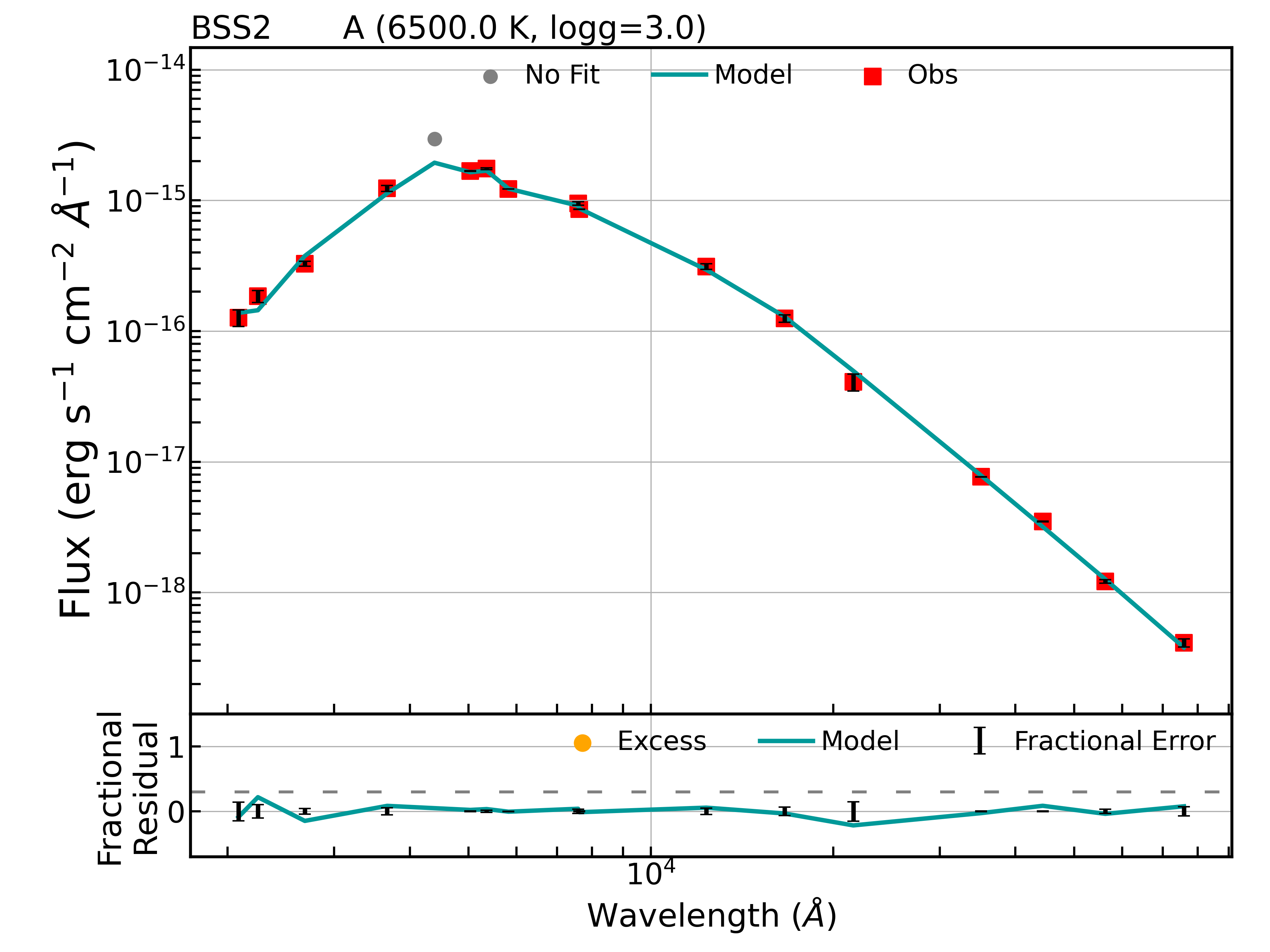}
		\caption*{}
	\end{subfigure}
	\quad 
	\begin{subfigure}[b]{0.48\textwidth}
   		\includegraphics[width=1.0\textwidth]{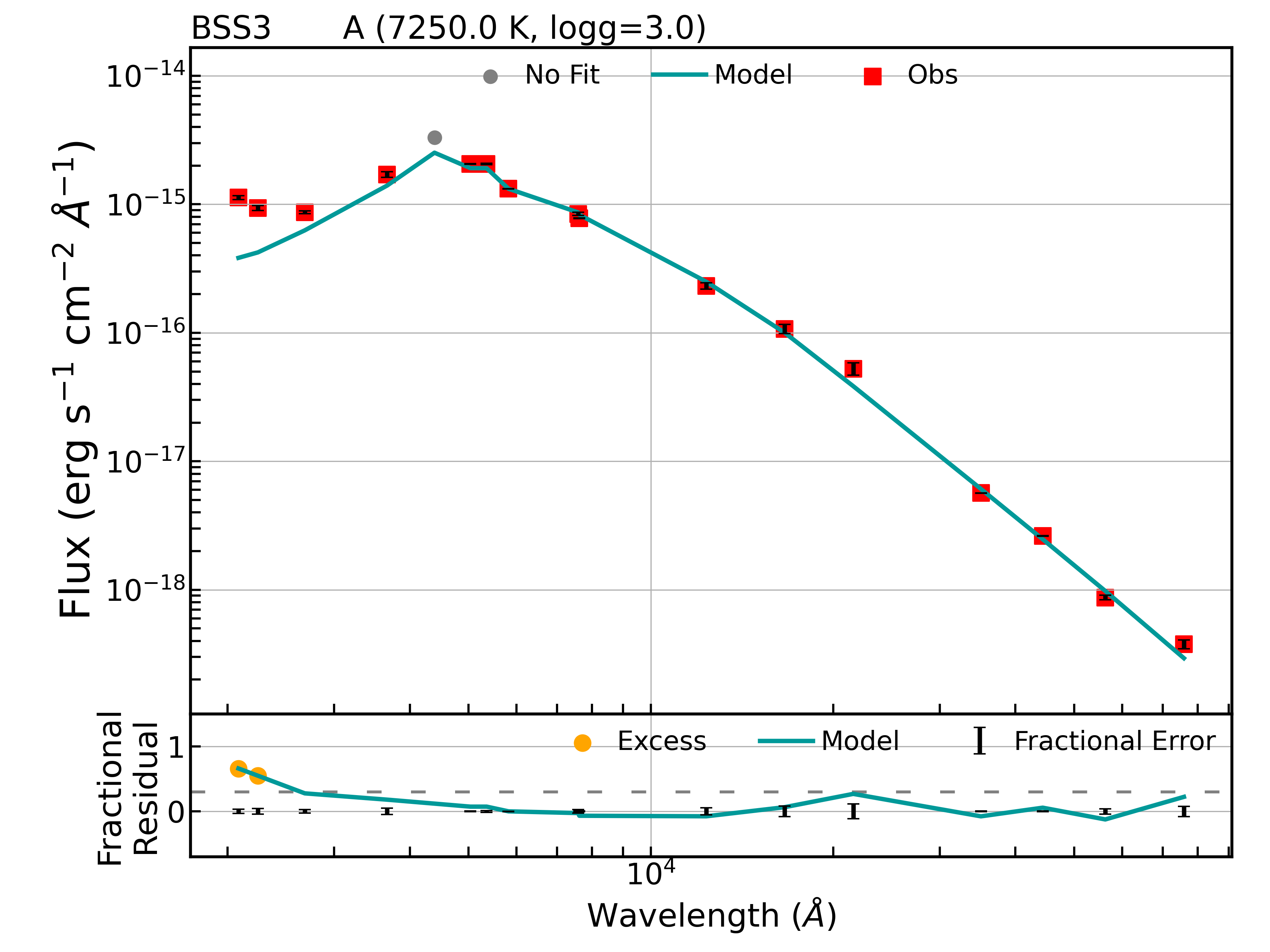}
		\caption*{}
	\end{subfigure}
	\quad 
	\begin{subfigure}[b]{0.48\textwidth}
		\includegraphics[width=1.0\textwidth]{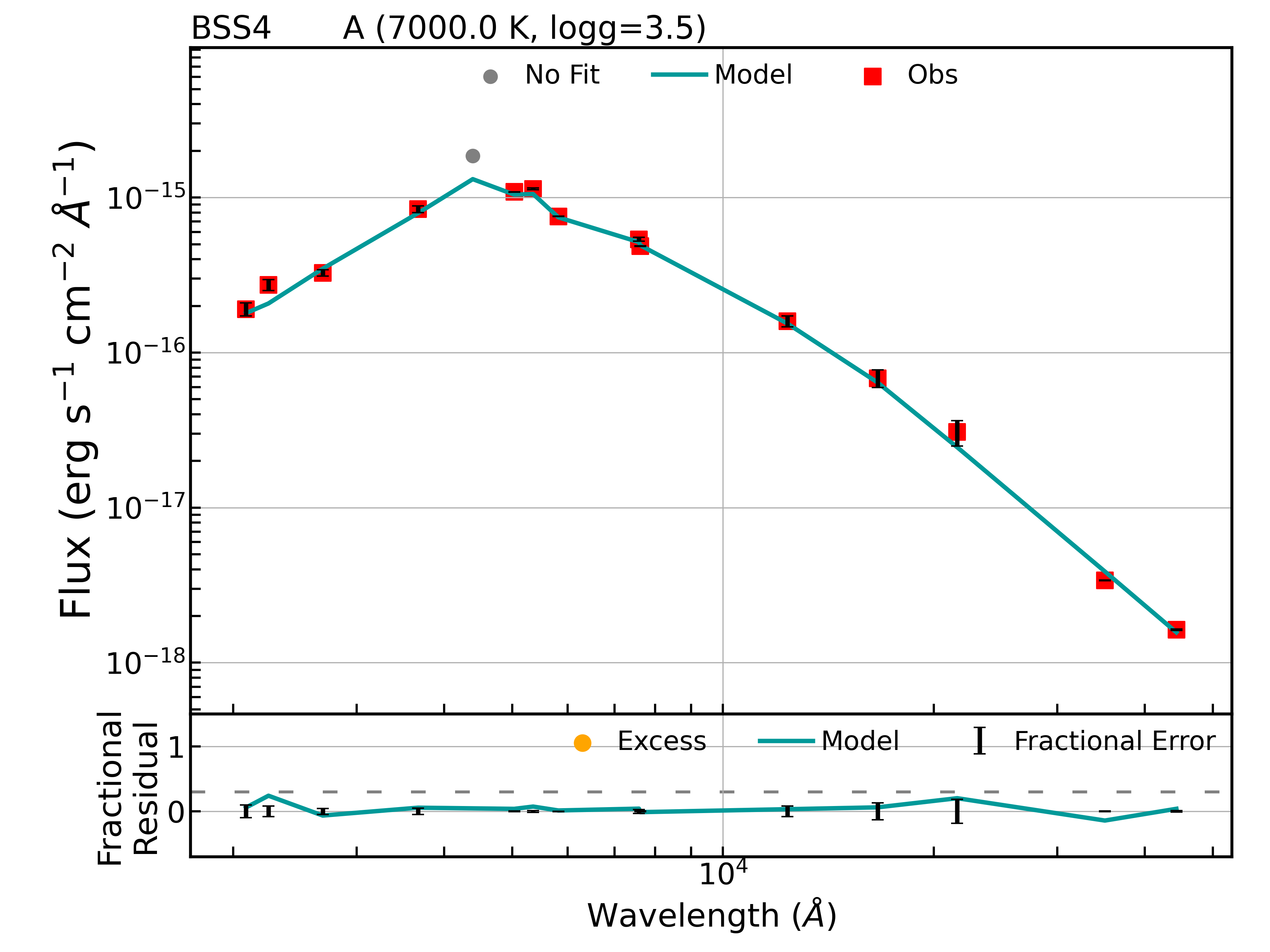}
		\caption*{}
	\end{subfigure}
	\quad 
	\begin{subfigure}[b]{0.48\textwidth}
   		\includegraphics[width=1.0\textwidth]{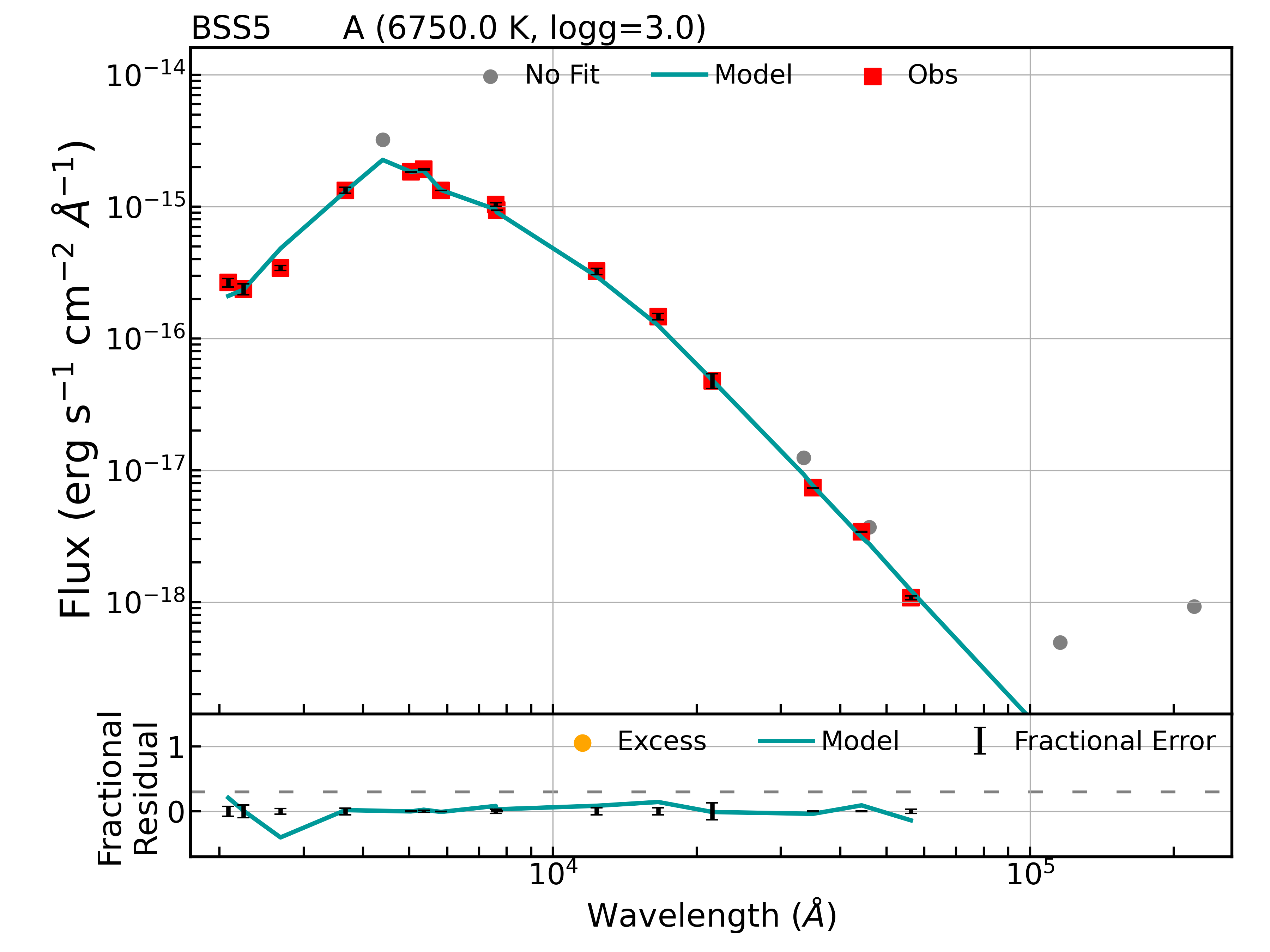}
		\caption*{}
	\end{subfigure}
	\quad
	\begin{subfigure}[b]{0.48\textwidth}
    		\includegraphics[width=1.0\textwidth]{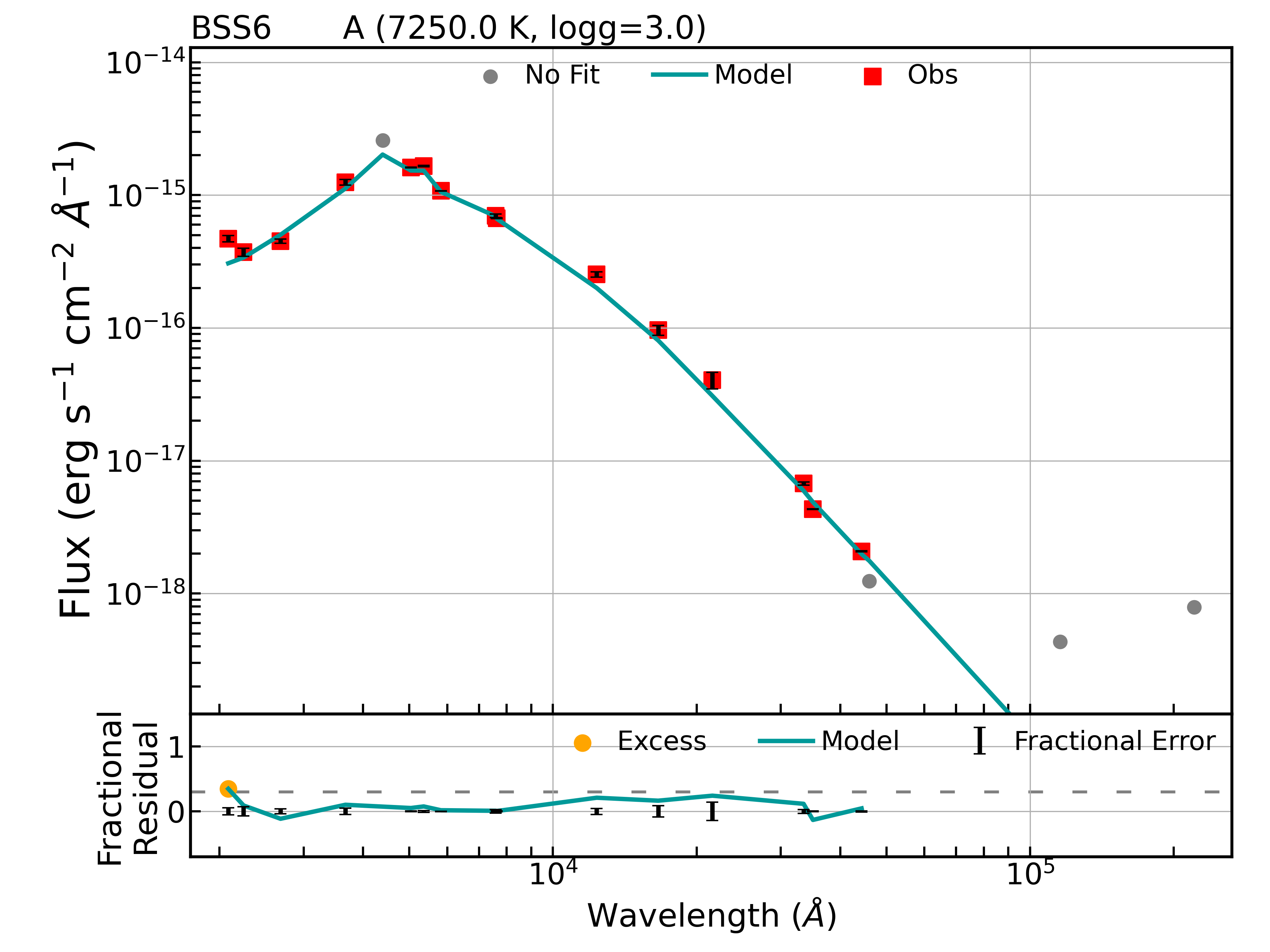}
		\caption*{}
	\end{subfigure}
	\caption{The single component fit SEDs for BSS. The upper panels show single SEDs fitted to extinction corrected fluxes with Kurucz stellar models. The red squares with black error bars show extinction corrected observed fluxes and errors in them. The blue curves show the fitted Kurucz stellar models. The lower panels depict the fractional residual at each data point, while the grey dashed lines represent the expected threshold value for measuring excess in flux values.}
	\label{fig:BSS_SED1}
\end{figure*}
\begin{figure*}
    \ContinuedFloat
    \centering
    \begin{subfigure}[b]{0.48\textwidth}
    		\includegraphics[width=1.0\textwidth]{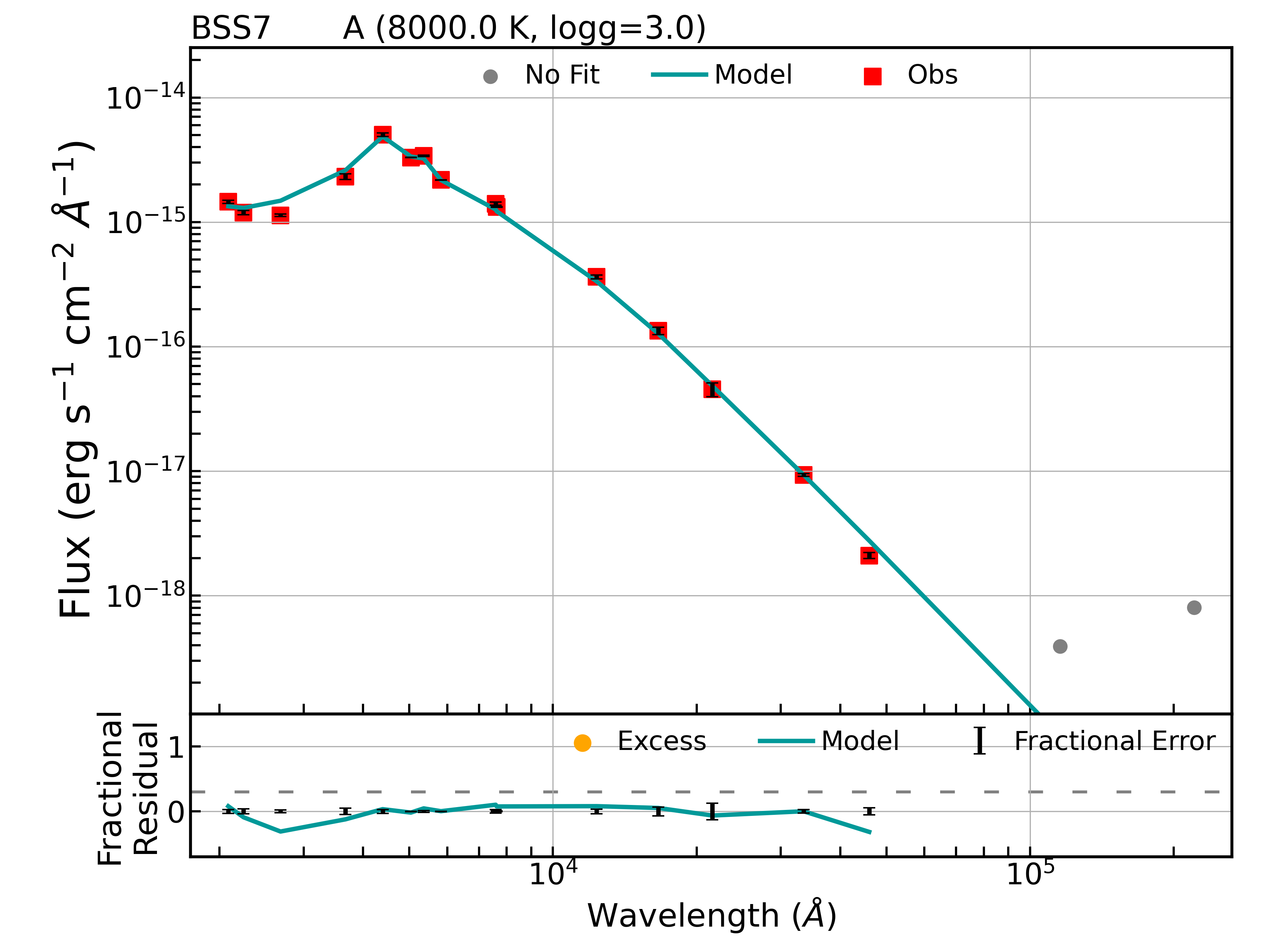}
		\caption*{}
	\end{subfigure}
	\quad
	\begin{subfigure}[b]{0.48\textwidth}
		\includegraphics[width=1.0\textwidth]{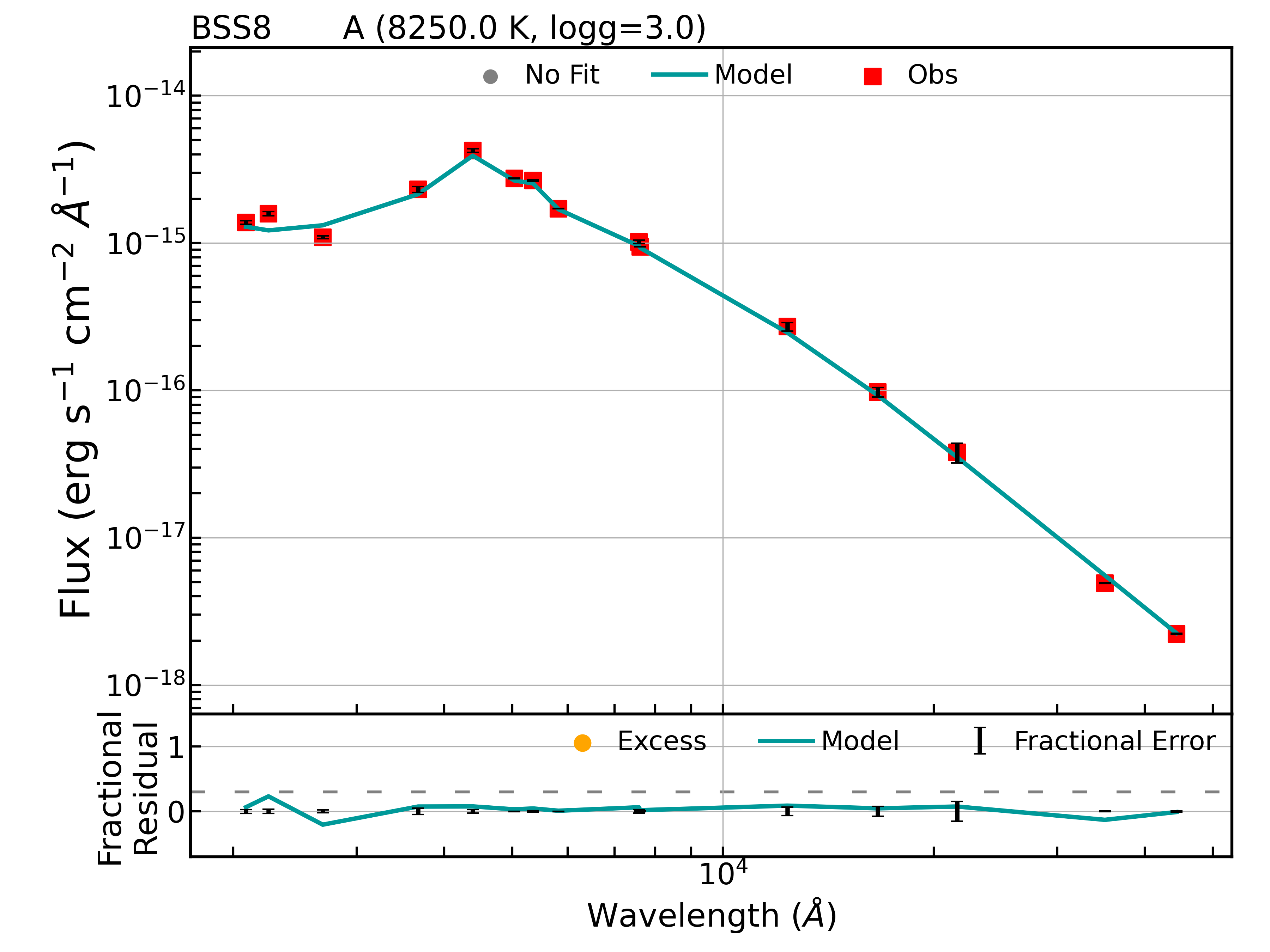}
		\caption*{}
	\end{subfigure}
	\quad
	\begin{subfigure}[b]{0.48\textwidth}
    		\includegraphics[width=1.0\textwidth]{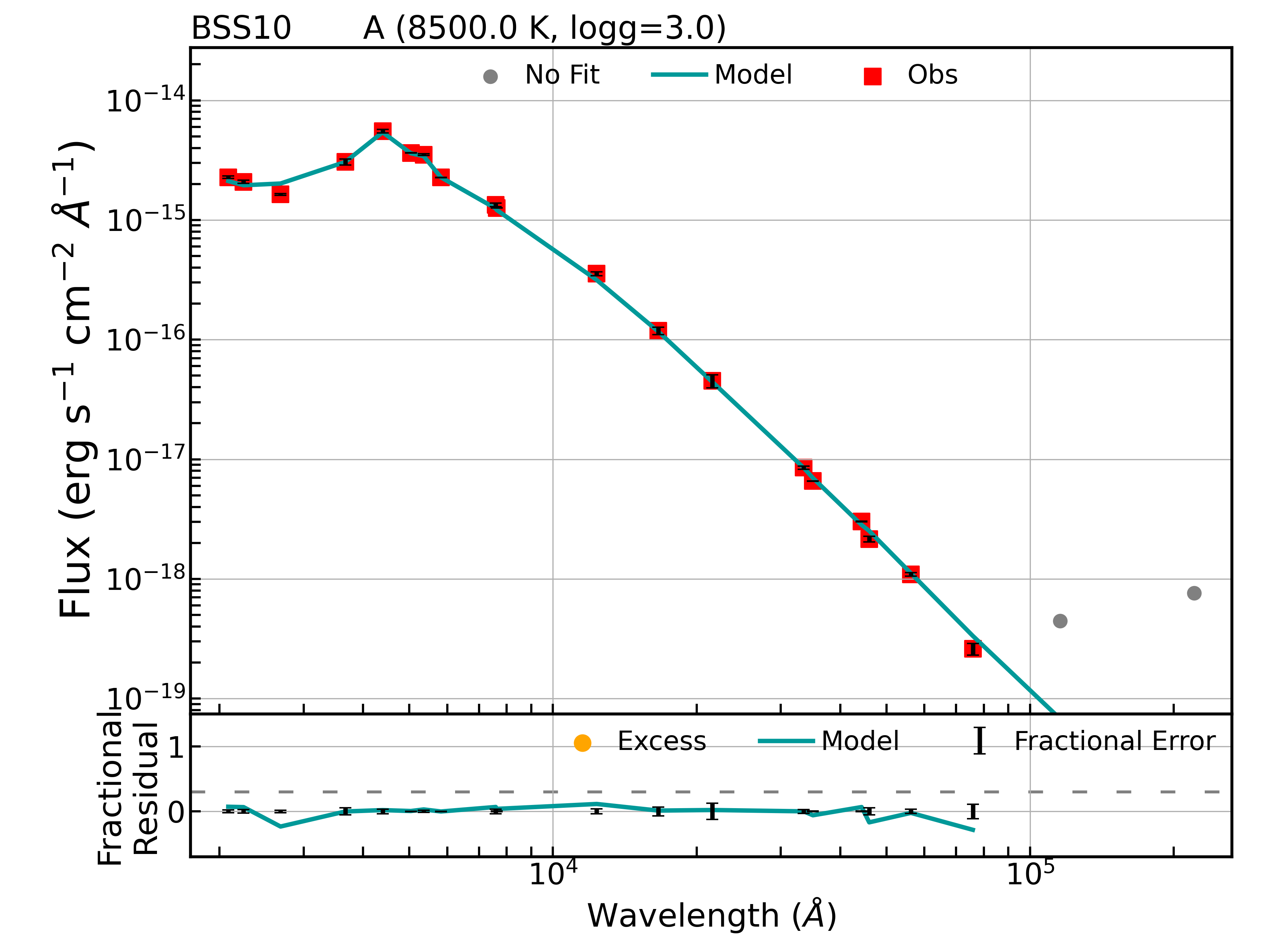}
		\caption*{}
	\end{subfigure}
	\quad 
	\begin{subfigure}[b]{0.48\textwidth}
   		\includegraphics[width=1.0\textwidth]{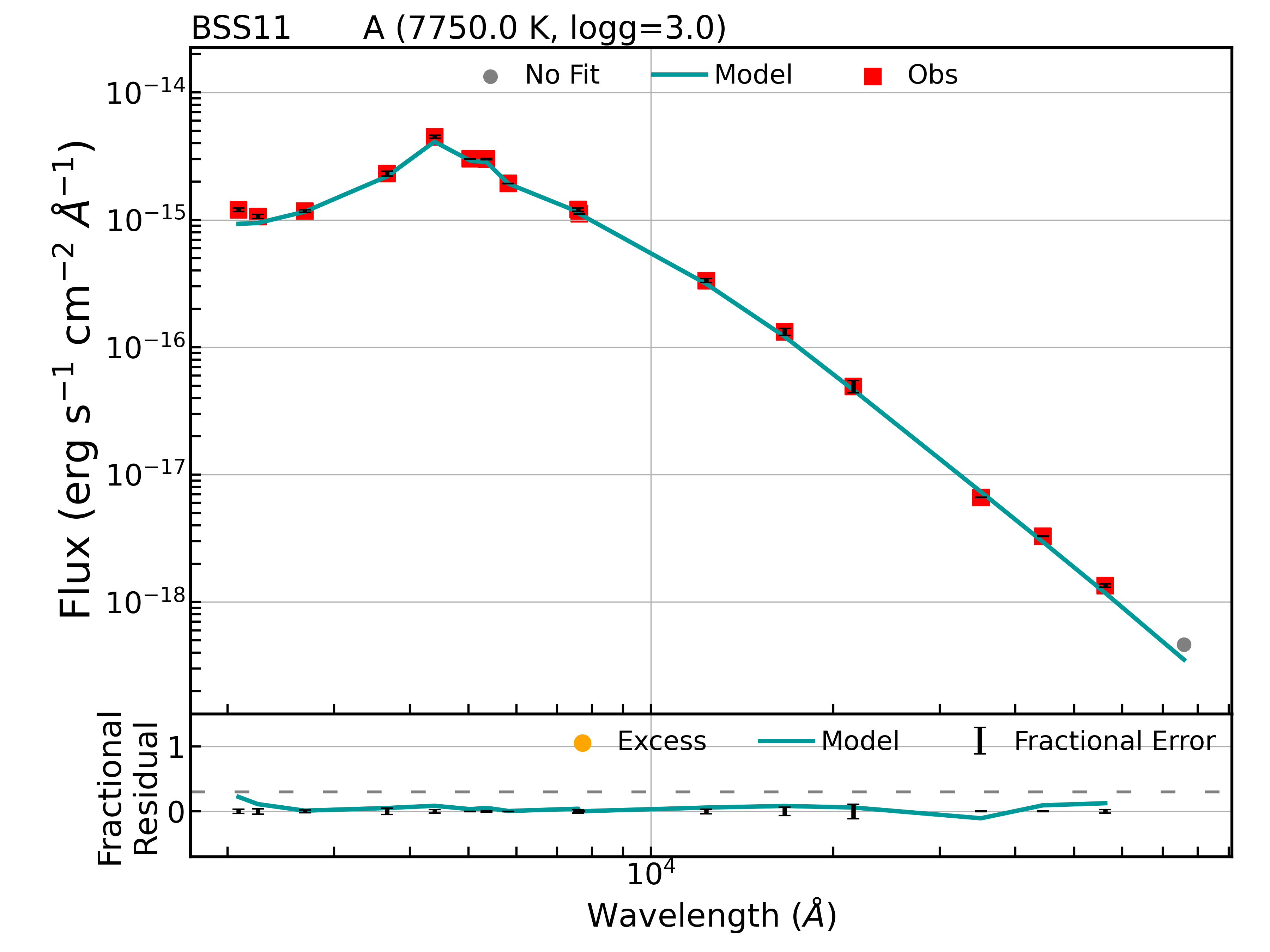}
		\caption*{}
	\end{subfigure}
	\quad 
	\begin{subfigure}[b]{0.48\textwidth}
   		\includegraphics[width=1.0\textwidth]{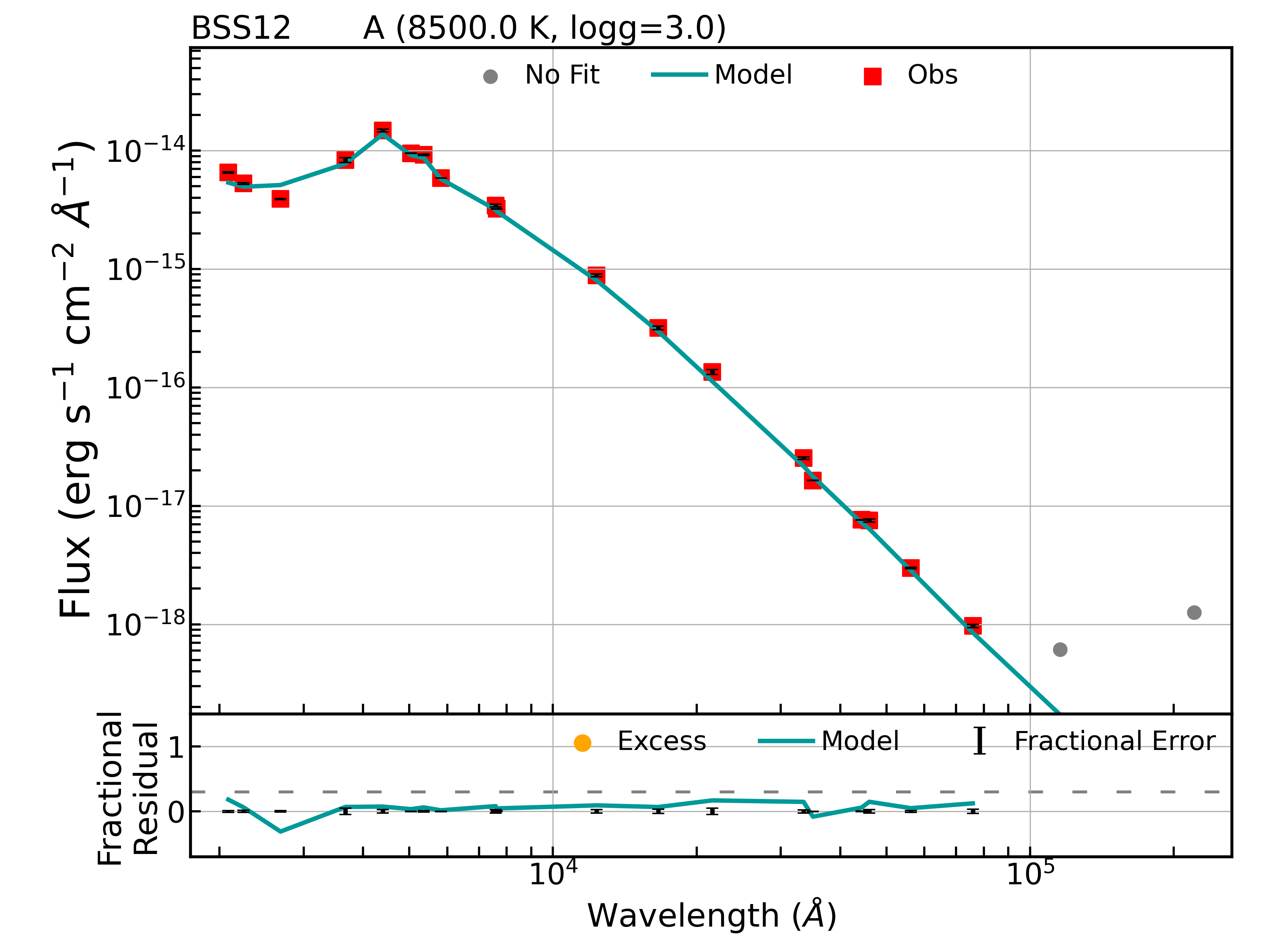}
		\caption*{}
	\end{subfigure}
	\caption{The single component fit SEDs for BSS.}
\end{figure*}
\begin{figure*}
	\includegraphics[width=0.8\textwidth]{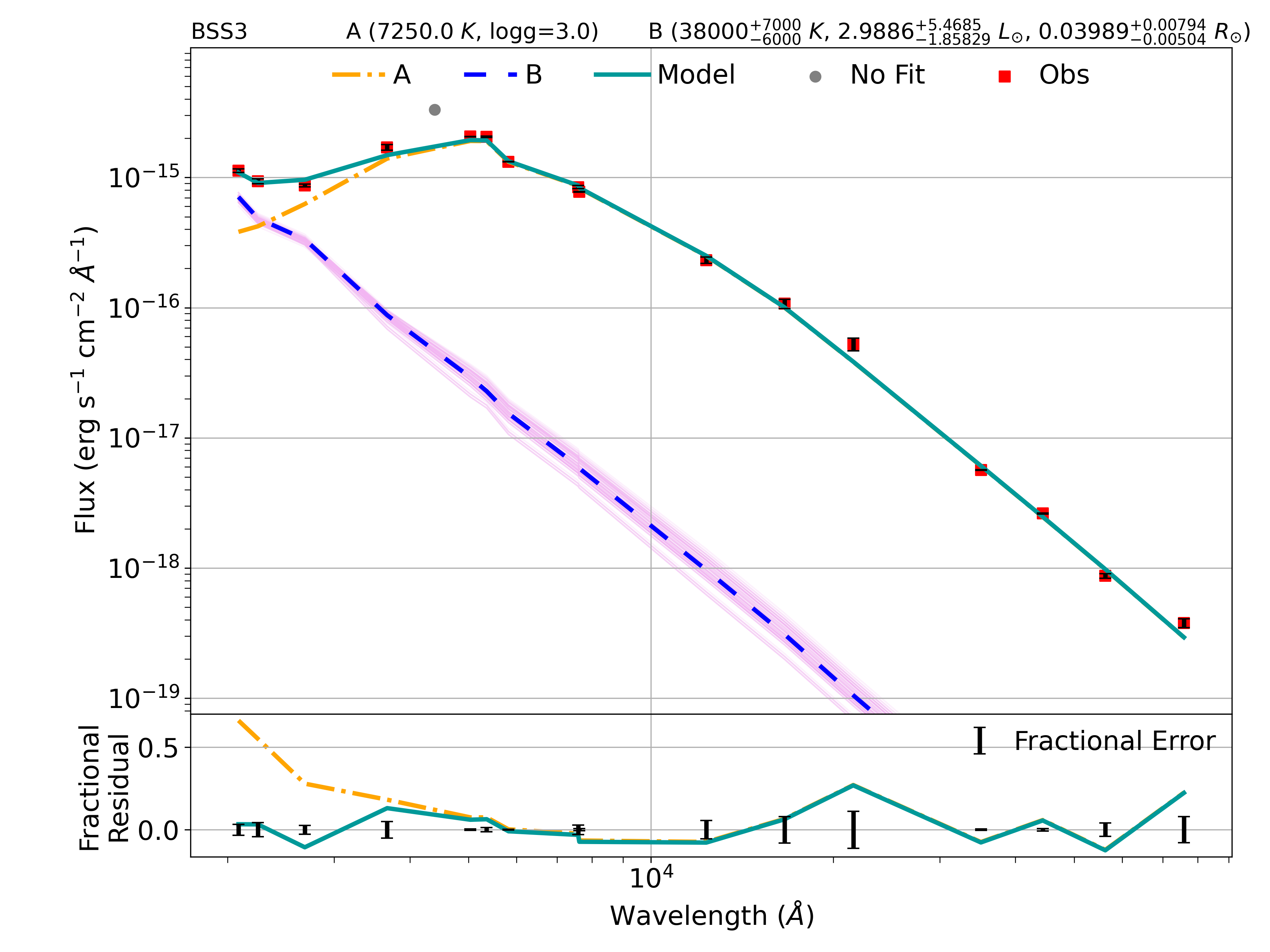}
    \caption{The double component fit SED for BSS3. The upper panel shows double component SED of BSS3, where the extinction corrected fluxes and errors in them are shown as red squares with black error bars, orange dashed dotted line shows SED of cool (A) component, blue dashed line shows SED of hot (B) component with 100 iterations of SEDs shown as pink solid lines. The data points not included for SED fitting are shown as gray dots. The lower panel shows fractional residual where the green solid line and yellow dashed dotted line is same as the upper panel at each datapoint.}
    	\label{fig:Double_SED}
\end{figure*}
\begin{figure}
   	\includegraphics[width=0.48\textwidth]{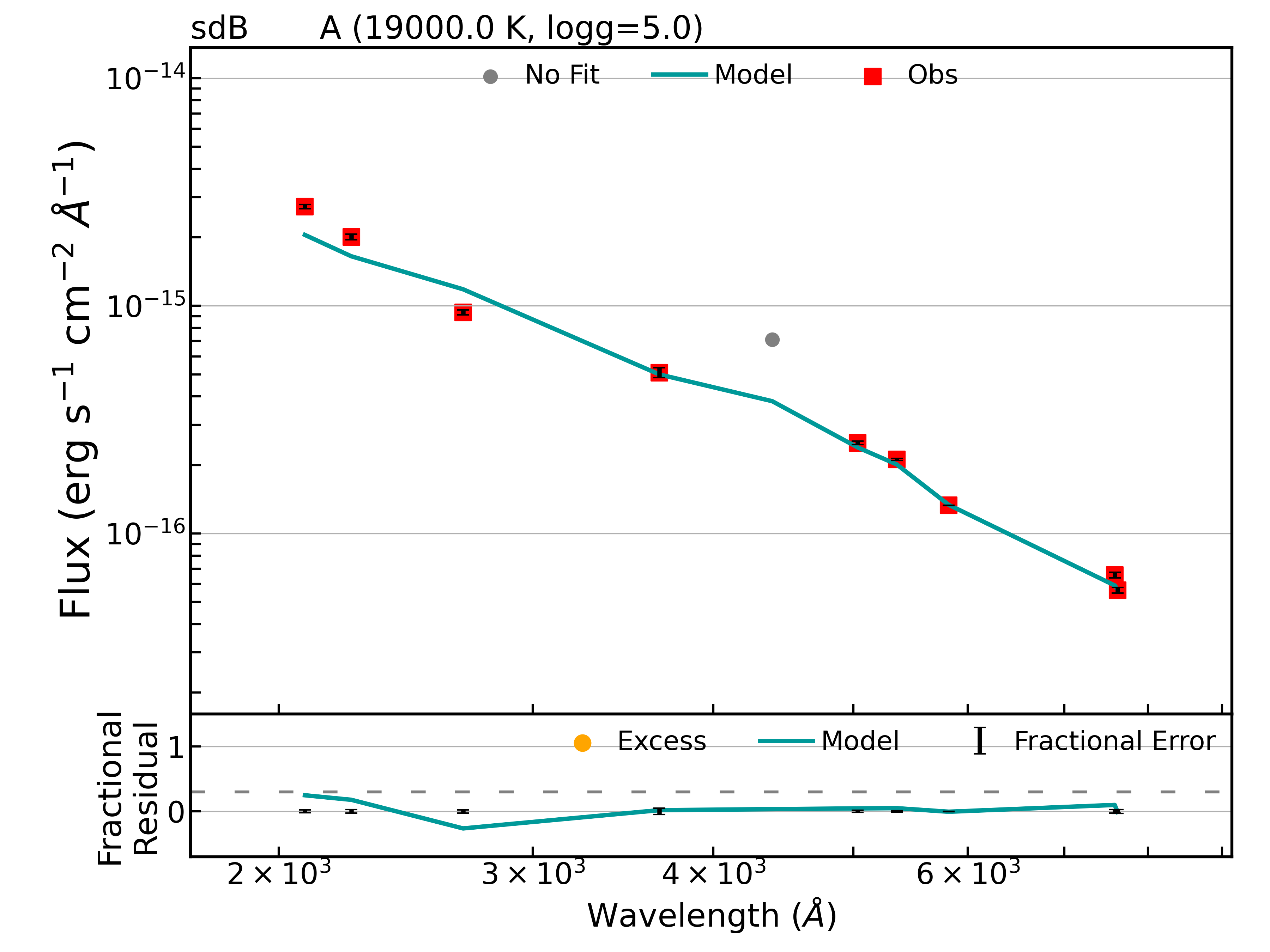}
	\caption{A single component SED of the sdB candidate. The details of the figure are the same as Figure \ref{fig:BSS_SED1}.}
	\label{fig:sdB_SED}
\end{figure}
\begin{figure*}
	\includegraphics[width=0.5\textwidth]{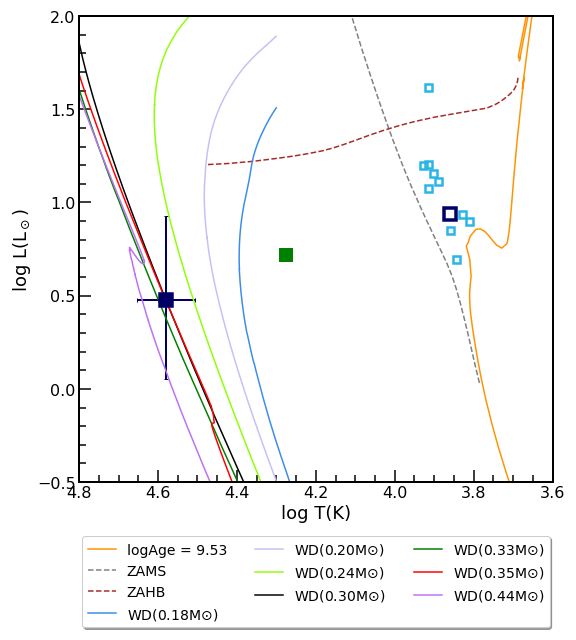}
    \caption{The H-R diagram of log(age) = 9.53 and Z = 0.01 shown as orange solid curve, ZAMS shown as grey dashed line, ZAHB is shown as the brown dashed curve. The low-mass WD models from \citep{Panei2007} are shown as solid curves. The BSS candidates with single SED fits are shown as light blue open squares and green filled square shows the sdB candidate. The hot component of  BSS3 is shown as blue filled square lying on the low-mass WD models, whereas the cooler component of BSS3 is shown as a blue open square above the main sequence turnoff.}
    	\label{fig:HRD}
\end{figure*}
\section{The Color magnitude diagram}
\label{sec:CMD}
The CMD of the cluster members identified using the ML-MOC algorithm on \textit{Gaia} EDR3 data is presented in Figure \ref{fig:cmd} with the plotted PARSEC isochrone \citep{Bressan2012}. The cluster members are depicted as grey dots, and their counterparts in UVW1 filter of Swift/UVOT within 1$\arcsec$ tolerance are depicted as orange dots. The PARSEC isochrone is plotted using the known cluster parameters such as age = 3.4 Gyr, Z = 0.01, distance = 4810 pc, and A$_{\mathrm{V}}$ = 0.43 taken from \citet{Vaidya2020}. The extinction (A$_{\mathrm{V}}$) we use to plot the isochrone is consistent within the errors to the extinction value estimated by \citet{Carraro2014} as A$_{\mathrm{V}}$ = 0.465$\pm$0.062. We identify 14 BSS candidates in this cluster using the criteria adopted by \citet{Rao2021}, which are shown as blue open circles in Figure \ref{fig:cmd}. The BSS population of this cluster has been previously identified by \citet{Vaidya2020} and \citet{Rain2021} using the Gaia DR2 data. Both identified 14 BSS candidates in this cluster. Our 13 BSS candidates are common with \citet{Rain2021}, however, one BSS candidate from \citet{Rain2021} is at $\sim$16.95$\arcmin$ from the cluster center, which is beyond our estimated cluster radius. Our 9 BSS candidates are common with \citet{Vaidya2020}, however, the remaining 5 BSS candidates from \citet{Vaidya2020} are below the main sequence turnoff and are not BSS candidates according to \citet{Rao2021} BSS selection criteria. Our additional 5 BSS candidates are not \citet{Vaidya2020} members due to their stringent membership selection criteria. 

Among the 14 identified BSS candidates, 12 are detected by Swift/UVOT, and two are outside the FOV of Swift/UVOT (see left panel of Figure \ref{fig:cmd}). Of the remaining 12 BSS candidates, 6 are detected in all the four Spitzer/IRAC filters, one in each of I1, I2, and I3, three in just I1 and I2, and two in none of the Spitzer/IRAC filters. As can be seen in Figure \ref{fig:cmd}, one BSS candidate, BSS12 (Gaia EDR3$\_$ID = 5507236216326238336), is located above the zero-age horizontal branch (ZAHB, brown dashed line). In clusters with ages between 0.5 and 4.0 Gyr, only red clump stars are detected owing to the initialization of core helium burning before the core becomes degenerate. The intermediate age of Melotte 66 implies that BSS12 is unlikely to be a blue horizontal branch candidate. Therefore, we consider it as a BSS candidate and do the further analysis accordingly. We identified 2 yellow straggler stars (YSS) candidates shown as blue open squares and one sdB candidate located at G = 18.6 mag located next to the extreme horizontal branch. This sdB candidate is identified in all the three filters of Swift/UVOT, and is also identified as a cluster member by \citet{Carraro2014}. Of the 12 BSS candidates, BSS9 has a bright source within 3$\arcmin$ radius from it. Therefore, in the present work, we study the remaining 11 BSS candidates and the sdB candidate, which have been detected in all three filters of Swift/UVOT.
\section{Results and Discussion}
\label{sec:SEDs}
\subsection{Spectral Energy Distributions}
We aim to estimate fundamental parameters such as the temperature, luminosity, and radius of the 12 BSS candidates and the sdB candidate of Melotte 66 by constructing their SEDs. For the present work, we use multiwavelength data ranging from NUV to IR, where we take NUV data from Swift/UVOT \citep{Roming2005}, optical data from \textit{Gaia} EDR3 \citep{Gaiaedr32021} and \citet{Carraro2014}, infrared data from Two Micron All Sky Survey \citep[2MASS;][]{Cohen2003}, Spitzer/IRAC \citep{Fazio2004}, and Wide-field Infrared Sky Explorer \citep[WISE;][]{Wright2010}. In order to generate SEDs, we make use of an online tool, Virtual Observatory SED Analyzer\footnote{\url{http://svo2.cab.inta-csic.es/theory/vosa/index.php}} \citep[VOSA;][]{Bayo2008}. The steps to create SEDs are as follows:
\begin{enumerate}
    \item We initially provide VOSA with source coordinates, A$_{\mathrm{V}}$ = 0.43, distance = 4810 pc, and fluxes of the cluster sources from Swift/UVOT, Spitzer/IRAC, and \citet{Carraro2014}. VOSA assembles fluxes in optical and infrared wavelength bands from the aforementioned archival databases within our selected 1.5$\arcsec$ tolerance. The value of A$_{\mathrm{V}}$ = 0.43 provided to VOSA was used to fit isochrone to the CMD (see Figure \ref{fig:cmd}). Using the extinction law given by \citet{Fitzpatrick1999} and \citet{Indebetouw2005}, VOSA corrects all of the observed fluxes for extinction. The extinction-corrected fluxes are then utilized to construct SEDs.
    \item  We fit the SEDs with the Kurucz stellar model \citep{Castelli1997}. We first fit the model SED to just optical and IR data points, omitting UV data points in order to investigate the source's true flux at UV wavelengths. In order to fit the model to SEDs, we keep temperature and log \textit{g} as free parameters while keep a range of metallicity from $-$0.5 to 0.0 as the cluster metallicity is known to range from [Fe/H] = $-$0.33 -- $-$0.17. The reduced chi-square ($\chi_r$) minimization approach is then used to get the best fitted SED. The value of the $\chi_r^2$ is calculated using the following formula:
    \begin{equation} \chi^2_r = \frac{1}{N-n_p}\sum_{i=0}^{N} \frac{(F_{o,i}-M_d F_{m,i})}{\sigma_{o,i}^2}
    \end{equation}
    where N is the total number of photometric data points, $n_{\mathrm{p}}$ is the total number of fitted model parameters, $F_{o,i}$ is the observed flux, $F_{m,i}$ is the theoretical flux predicted by the model, $\sigma_{o,i}$ is the observational error in the flux, and $M_d$ is the multiplicative dilution factor given as (R/D)$^2$, R is the radius of an object and D is the distance between object and the observer. Based on the lowest $\chi_r^2$ values, VOSA produces a few best-fitted SEDs. Even when the SED fits look good visually, sometimes the corresponding $\chi_r^2$ values are very large. This is caused by very small errors in one or more observed fluxes. In that scenario, even a small deviation of model flux from the observed flux leads to a large $\chi_r^2$ value. To address this issue, a new parameter called "visual goodness of fit", abbreviated as \textit{vgf$_b$}, is also calculated by VOSA. This is a modified $\chi_r^2$ in which the flux errors less than 0.1$\times$flux are replaced by 0.1$\times$flux. SEDs with \textit{vgf$_b$} smaller than 10 -- 15 are deemed to be good fits \citep{Rebassa2019,Rebassa2021,Jim2019}. Based on the best SED fit, VOSA provides the estimated values of parameters and their errors where the errors are basically the half of the bin size.
    \item The nature of the object at UV wavelengths is characterized after fitting the optical and IR data points. Sources with excess in UV fluxes are expected to either have hot spots, coronal or chromospheric activities or unresolved hot companions. Since no X-ray data is currently available for Melotte 66, more data is needed to understand if the excess is either due to hot spots, coronal or chromospheric activities. We fit sources with greater than 30$\%$ excess in more than one UV filter with double component SEDs using the python code \textit{Binary$\_$SED$\_$Fitting}\footnote{\url{https://github.com/jikrant3/Binary_SED_Fitting}}. We briefly explain the method here, while for the details of the method, readers are referred to \citet{Jadhav2021}. In order to fit a second component SED, we use Koester WD model \citep{Koester2010} that has a range in temperature of 5000 -- 80000 K and in log \textit{g} of 6.5 -- 9.0. For this, we keep temperature and log \textit{g} as free parameters while restrict metallicity from $-$0.5 to 0.0. To further determine the parameters of the hotter components and errors in those estimated parameters, we use the statistical approach adopted by \citet{Jadhav2021}. For this, we generate 100 SEDs by adding Gaussian noise proportional to the errors to each observed fluxes. We then fit those 100 SEDs with double component SEDs and estimate parameters of the hotter components for each of the 100 double fit SEDs. The median and the standard deviation of those 100 values of each parameter serve as the parameter value and the associated error in it, respectively. 
\end{enumerate}
\subsubsection{SEDs of the BSS candidates}
\textit{BSS with single-component SED fit:}
We constructed SEDs for 11 BSS candidates using the multiwavelength data ranging from NUV to IR as described earlier. We used a total of 14 -- 20 data points to build the SEDs. The data points which either have bad photometry detected by VOSA or deviates much from the model flux are not included to construct SEDs. For some of the BSS candidates, only upper limit in W3 and W4 filters of WISE catalog are available, therefore we have excluded them from the fitting. The excluded data points are shown as grey dots in Figure \ref{fig:BSS_SED1}. The spatial coordinates, Gaia IDs, and Swift/UVOT fluxes are cataloged in Table \ref{tab:Table_flux_bss}. As shown in Figure \ref{fig:BSS_SED1}, 9 out of 11 BSS candidates are successfully fitted with single SEDs. BSS3 has fractional residual greater than 0.3 in UVW2 and UVM2 filters, therefore we consider it for a double SED fit. In BSS6, only one filter, UVW2, has fractional residual greater than 0.3. The double component fit, on the other hand, returned a wide temperature range of the hotter component. Therefore, we fitted a single component SED to BSS6.

From the SED fits, the temperatures of the BSS candidates range from 6500 K to 8500 K, and the luminosities range from 4.90  L$_{\sun}$ to 41.70 L$_{\sun}$. BSS12 has a temperature of 8500 K but has a highest luminosity of 41.79 L$_{\sun}$. The fundamental parameters such as temperature, log \textit{g}, luminosity, radius, $\chi_2^2$, and \textit{vgf$_b$} estimated by single component SED fitting using VOSA are listed in Table \ref{tab:Table_bss_params}. When the single-mass isochrone fitted to the main sequence is extrapolated, the masses of the BSS candidates range from 1.29 to 4.15 M$_{\sun}$. As the turnoff mass of the cluster is 1.17 M$_{\sun}$, the BSS progenitors have gained mass at least from 0.12 M$_{\sun}$ to 2.98 M$_{\sun}$. On that note, we have 1 BSS candidate with mass 0.0 -- 0.2 M$_{\sun}$ greater than the turnoff mass, 6 BSS candidates with masses 0.2 -- 0.5  M$_{\sun}$ greater than the turnoff mass, 6 BSS candidates with masses 0.5 -- 1.0  M$_{\sun}$ greater than the turnoff mass, 1 BSS candidate with mass >1.0 M$_{\sun}$ than the turnoff mass \citep{Leiner2021}. BSS12 has a mass of 4.15 M$_{\sun}$ that is 2.98 M$_{\sun}$ much massive than the turnoff mass of the cluster. Such a massive BSS can form either because of the multiple mass transfer events or merging of more than 2 main sequence turnoff stars \citep{Jadhav2021b}.

\textit{BSS with two-component SED fit:}
We have detected excess flux in two filters of Swift/UVOT, UVW2 and UVM2 for BSS3, therefore we have fitted a double component SED. The cooler component of BSS3 has a temperature of 8250 K and a luminosity of 8.67$\pm$0.67 L$_{\sun}$. The single component SED of BSS3 has $\chi_r^2$ = 116.1 and \textit{vgf$_b$} = 7.73 which get reduced to $\chi_r^2$ = 60.96 and \textit{vgf$_b$} = 1.91 for the composite fit. The hotter component of BSS3 has a temperature of 38000$_{-6000}^{+7000}$ K, a luminosity of 2.99$_{-1.86}^{+5.47}$ L$_{\sun}$, and a radius of 0.04$_{-0.005}^{+0.008}$ R$_{\sun}$. The parameters obtained indicate that the hotter component is likely to be low mass WD. Figure \ref{fig:Double_SED} shows the composite SED fit for BSS3, whereas the BSS fluxes and obtained parameters are listed in Table \ref{tab:Table_flux_bss} and Table \ref{tab:Table_bss_params}, respectively. 

Figure \ref{fig:HRD} represents the H-R diagram of the Melotte 66 open cluster. The PARSEC isochrone of log(age) = 9.53 and Z = 0.01 is shown as orange solid line, zero-age main sequence (ZAMS) is shown as grey dashed line, ZAHB is shown as the brown dashed line. On the H-R diagram, we have also shown helium WD models \citep{Panei2007} of masses ranging from 0.18 M$_{\sun}$ to 0.44 M$_{\sun}$. The cooler component of BSS3 is shown as blue open square along the BSS sequence above the main sequence turnoff and its hot companion is shown as blue filled square with error bars lying on the helium WD models. According to the helium WD models, it has mass of 0.24 -- 0.44 M$_{\sun}$ and age of 19.95 -- 38.72 Myr. This indicates that BSS3 is formed via either the Case A or Case B mass-transfer channel.

\subsubsection{SED of the sdB candidate}
We identify one sdB candidate in this cluster that has membership probability of 0.94. This sdB candidate was also detected by Swift/UVOT survey as well as it is in the catalog for cluster members provided by \citet{Carraro2014}. In the variability study, \citet{Zloczewski2007} mentioned it as a non-variable sdB star. They further reported its M$_\mathrm{V}$ = 5.4 mag, which happens to be around the fainter end of absolute magnitude distribution for hot subdwarfs in the field. As shown in Figure \ref{fig:sdB_SED}, the sdB candidate is successfully fitted with a single component SED. We obtain its temperature as 19000 K, luminosity as 5.29$\pm$0.45 L$_{\sun}$, radius as 0.21$\pm$0.009 R$_{\sun}$. The sdB candidate is depicted as green filled square right to the 0.20 M$_{\sun}$ helium WD model on the H-R diagram shown in Figure \ref{fig:HRD}.   

\subsection{Variable BSS according to TESS data}
\begin{figure}
   		\includegraphics[width=0.48\textwidth]{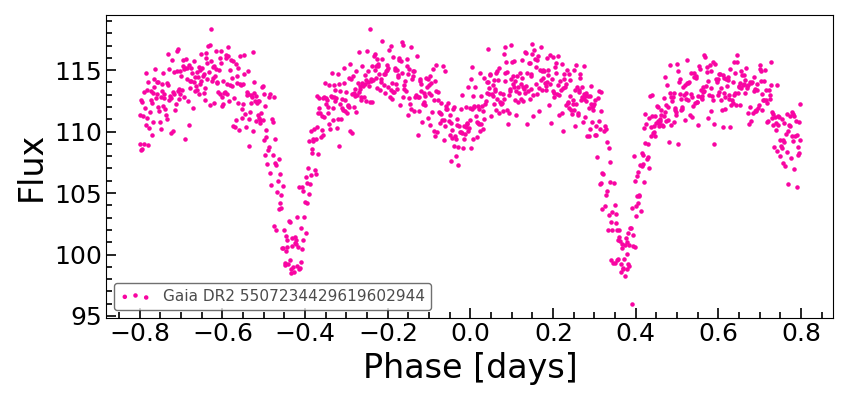}
	\caption{The TESS light curve of BSS6 taken from \citet{Nardiello2020}.}
	\label{fig:light_curves}
\end{figure}
For all our BSS candidates, except BSS9, which has a nearby source within 3$\arcsec$ from it, we searched the TESS light curves in \citet{Nardiello2020}. We found one BSS candidate, BSS6, as a variable star.

\textit{BSS6:} \citet{Zloczewski2007} identified it as an eclipsing binary with a period of 0.8015 days. They examined the variability of several stars of the Melotte 66 open cluster using BVI photometric data obtained from the 1-m Swope telescope at Las Campanas Observatory. We also identified BSS6 as an eclipsing binary using the light curve obtained from \citet{Nardiello2020}. The  light curve of BSS6 is shown in Figure \ref{fig:light_curves}. We computed the period as 0.8006 days using Lomb-scargle periodogram \citep{Press1989}. Our estimated period is quite similar to that of \citet{Zloczewski2007}. As previously stated, BSS6 has a residual in the UVW2 filter greater than 0.3. However, we were unable to get a reliable double component SED fit as the temperature of the hot companion was reaching up to the maximum limit of the Koester model, i.e., 80000 K. The detected excess could be owing to an associated hot companion or magnetic activities such as flares or spots. BSS6 is listed as an EA type eclipsing binary (beta Persei or Algol-type eclipsing binary) by \citet{Kazarovets2011} and \textit{Gaia} DR3 variability catalog \citep{Eyer2022}. The $\beta$ Persei-type (Algol) binaries are eclipsing systems in which the donor has transferred enough of its mass to the accretor that it has now remained as a less massive component while still filling its Roche Lobe \citep{Peters2001}. Additional spectroscopic observations or light curve modelling are required to constrain its spectroscopic parameters, rotational velocity, mass, etc. Nevertheless, such a short period of the eclipsing binary supports the formation of BSS6 via the Case A mass-transfer channel.
\begin{table*}
	\centering
	\caption{For each of star studied in the present work, coordinates in columns 2 and 3, Gaia source ID in column 4, Swift/UVOT UVW2, UVM2, UVW1 fluxes and their associated errors are in columns 5, 6, and 7, respectively. All values of fluxes are given in ergs $s^{-1}$ $cm^{-2}$ $\si{\angstrom}^{-1}$.}
	\label{tab:Table_flux_bss}
	\begin{tabular}{ccccccc}
		\hline
		\\
		Name & RA & DEC & Gaia EDR3 source$\_$id & UVOT.UVW2$\pm$err & UVOT.UVM2$\pm$err & UVOT.UVW1$\pm$err  \\
		\\
		\hline
		\\
		sdB & 111.57540 & $-$47.74427 & 5507046035169605120 & 8.02e-16$\pm$1.71e-17 & 6.08e-16$\pm$1.74e-17 & 4.15e-16$\pm$1.02e-17 \\
		BSS1 & 111.57101 & $-$47.68414 & 5507234086022210688 & 4.89e-16$\pm$1.35e-17 & 5.24e-16$\pm$1.63e-17 & 6.13e-16$\pm$1.23e-17 \\
		BSS2 & 111.55695 & $-$47.68872 & 5507234086022206208 & 3.73e-17$\pm$5.38e-18 & 5.59e-17$\pm$5.80e-18 & 1.45e-16$\pm$6.37e-18 \\
		BSS3 & 111.58645 & $-$47.69622 & 5507234395259864448 & 3.32e-16$\pm$1.11e-17 & 2.84e-16$\pm$1.22e-17 & 3.85e-16$\pm$1.01e-17 \\
		BSS4 & 111.60269 & $-$47.68522 & 5507234429619604224 & 5.62e-17$\pm$5.39e-18  & 8.27e-17$\pm$6.79e-18 & 1.45e-16 $\pm$6.7e-18 \\
		BSS5 & 111.58186 & $-$47.70574 & 5507046344410918528 & 7.82e-17$\pm$5.83e-18 & 7.17e-17$\pm$6.88e-18 & 1.52e-16$\pm$6.60e-18 \\
		BSS6 & 111.61125 & $-$47.67402 & 5507234429619602944 & 1.38e-16$\pm$7.54e-18 & 1.13e-16$\pm$7.83e-18 & 2.00e-16$\pm$7.39e-18 \\
		BSS7 & 111.60693 & $-$47.71108 & 5507046653648577664 & 4.28e-16$\pm$1.25e-17 & 3.60e-16$\pm$1.37e-17 & 5.02e-16$\pm$1.14e-17 \\
		BSS8 & 111.52107 & $-$47.65420 & 5507235151174063744 & 4.06e-16$\pm$1.20e-17 & 4.80e-16$\pm$1.57e-17 & 4.85e-16$\pm$1.12e-17 \\
		BSS9 & 111.60993 & $-$47.61481 & 5507235627911411456 & 2.15e-15$\pm$2.71e-17 & 4.80e-16$\pm$1.57e-17 & 1.46e-15$\pm$1.88e-17\\
		BSS10 & 111.43099 & $-$47.71431 & 5507211404296051712 & 6.70e-16$\pm$1.54e-17 & 6.31e-16$\pm$1.75e-17 & 7.24e-16$\pm$1.34e-17 \\
		BSS11 & 111.50080 & $-$47.59403 & 5507238449709052544 & 1.93e-15$\pm$2.58e-17 & 1.60e-15$\pm$2.78e-17 & 1.73e-15$\pm$2.06e-17 \\
		BSS12 & 111.71883 & $-$47.60413 & 5507236216326238336 & 3.53e-16$\pm$1.19e-17 & 3.22e-16$\pm$1.30e-17 & 5.20e-16$\pm$1.17e-17 \\
		\\
		\hline
	\end{tabular}
\end{table*}
\begin{table*}
	\centering
	\caption{The list of the estimated parameters of the BSS and sdB candidates using best fitted SEDs. Whether the single component or two component SED is satisfactory in Column 2, log \textit{g} in Column 3, temperature, radius, and luminosity in Columns 4--6, the scaling factor by which the model has to be multiplied to fit the data in Column 7, the number of data-points fitted in Column 8, the reduced $\chi_r^2$ in Column 9, and the modified reduced $\chi_r^2$, \textit{vgf$_b$} in Column 10.}
	\label{tab:Table_bss_params}
	\begin{tabular}{cccccccccc}
		\hline
		\\
		Name & Component & log \textit{g} & T$_{\textit{eff}}$ & R & L & Scaling Factor & N$_{\textit{fit}}$ & $\chi_r^2$ & \textit{vgf$_b$}  \\
		    &   &   & [K] & [R$_{\sun}$] & [L$_{\sun}$] &   &   &  &
		\\
		\hline
		\\
		sdB & single & 5 & 19000$\pm$125 & 0.21$\pm$0.009 & 5.29$\pm$0.45 & 9.48e-25 & 9 & 54.8 & 2.95 \\
		BSS1 & single & 3 & 8250$\pm$125 & 1.94$\pm$0.08 & 16.0$\pm$1.34 & 8.24e-23 & 20 & 78.9 & 1.05 \\
		BSS2 & single & 3 & 6500$\pm$125 & 2.22$\pm$0.09 & 7.98$\pm$0.69 & 1.08e-22 & 16 & 25.2 & 0.93 \\
		BSS3 & single & 3 & 7250$\pm$125 & 1.85$\pm$0.08 & 8.67$\pm$0.75 & 7.52e-23 & 17 & 116.3 & 1.91 \\
		     & double & 7.0 & 38000$^{+7000}_{-6000}$ & 0.040$^{+0.008}_{-0.005}$ & 2.99$^{+5.47}_{-1.86}$ & 6.61e-26 & -- & 60.9 & 1.43 \\
		BSS4 & single & 3.5 & 7000$\pm$125 & 1.50$\pm$0.06 & 4.97$\pm$0.44 & 4.94e-23 & 14 & 61.5 & 1.04 \\
		BSS5 & single & 3 & 6750$\pm$125 & 2.15$\pm$0.09 &  8.72$\pm$0.76 & 1.01e-22 & 17 & 37.6 & 1.94 \\
		BSS6 & single & 3 & 7250$\pm$125 & 1.66$\pm$0.07 & 7.13$\pm$0.62 & 6.04e-23 & 17 & 62.9 & 2.1 \\
		BSS7 & single & 3 & 8000$\pm$125 & 1.98$\pm$0.08 & 14.46$\pm$1.25 & 8.65e-23 & 17 & 34.3 & 1.84 \\
		BSS8 & single & 3 & 8250$\pm$125 & 1.67$\pm$0.07 & 11.91$\pm$1.03 & 6.1e-23 & 15 & 71.9 & 1.23 \\
		BSS10 & single & 3 & 8500$\pm$125 & 1.84$\pm$0.08 & 15.96$\pm$1.35 & 7.44e-23 & 21 & 33.8 & 1.04 \\
		BSS11 & single & 3 & 7750$\pm$125 & 1.97$\pm$0.08 & 13.07$\pm$1.1 & 8.55e-23 & 15 & 78.4 & 1.31 \\
	    BSS12 & single & 3 & 8500$\pm$125 & 2.95$\pm$0.12 & 41.79$\pm$3.58 & 1.89e-22 & 21 & 150.4 & 1.51\\
		\\
		\hline
	\end{tabular}
\end{table*}
\section{Final remarks}
\label{sec:Final_remarks}
For the first time, we characterize the BSS and sdB candidates of the intermediate-age open cluster Melotte 66 using the UV data. We summarize results as follows.
\begin{enumerate}
    \item Using the robust membership determination algorithm ML-MOC on \textit{Gaia} EDR3 data, we have identified 1162 members including 14 BSS candidates, 2 YSS candidates, and one sdB candidate in this cluster. In the present work, we focus on studying 11 out of 14 BSS candidates, and the sdB candidate by constructing their multiwavelength SEDs. The sdB candidate and 9 of the 11 BSS candidates are successfully fitted with single component SED fits.
    \item BSS3 was successfully fitted with a composite SED fit that resulted in the discovery of a low-mass helium WD as a companion. The mass and age of the WD companion reveal that BSS3 likely formed 19 -- 38 Myr back via either the Case A or Case B mass-transfer channel.
    \item BSS6 is an Algol-type eclipsing binary with a period of 0.8006 days. The period and nature of the eclipsing binary suggest that it might have formed via the Case A mass-transfer channel. 
    \item We suggest that two of the eleven BSS candidates in this cluster are formed via mass-transfer channel.
    \item We have identified one sdB candidate located at the end of the extreme horizontal branch in the CMD which is successfully fitted with a single component SED. The identified sdB candidate shows the capability of the ML-MOC algorithm in discovering exotic populations in open clusters.
\end{enumerate}
\section*{Acknowledgements}
We are thankful to the anonymous referee for valuable comments which helped in improving the manuscript. This work has made use of early third data release from the European Space Agency (ESA) mission {\it Gaia} (\url{https://www.cosmos.esa.int/gaia}), Gaia eDR3 \citep{Gaiaedr32021}, processed by the {\it Gaia} Data Processing and Analysis Consortium (DPAC, \url{https://www.cosmos.esa.int/web/gaia/dpac/consortium}). Funding for the DPAC has been provided by national institutions, in particular the institutions participating in the {\it Gaia} Multilateral Agreement. This research has made use of the VizieR catalog access tool, CDS, Strasbourg, France. This research made use of {\small ASTROPY}, a {\small PYTHON} package for astronomy \citep{Astropy2013}, {\small NUMPY} \citep{Harris2020}, {\small MATPLOTLIB} \citep{Hunter4160265}, {\small SCIPY} \citep{Virtanen2020}. This research also made use Astrophysics Data System (ADS) governed by NASA (\url{https://ui.adsabs.harvard.edu}).

\section*{Data Availability}
The data underlying this article are publicly available at \url{https://gea.esac.esa.int/archive}. The derived data generated in this research will be shared on reasonable request to the corresponding author.


\bibliographystyle{mnras}
\bibliography{References} 




\bsp	
\label{lastpage}
\end{document}